\documentclass[11pt]{article}
\usepackage[final]{acl}
\usepackage{times}
\usepackage{latexsym}

\usepackage[T1]{fontenc}
\usepackage{makecell}

\usepackage[utf8]{inputenc}
\usepackage{microtype}
\usepackage{inconsolata}
\usepackage{enumitem}
\usepackage{amsmath}

\usepackage{graphicx}

\title{Microsoft Academic Graph Information Retrieval for Research Recommendation and Assistance}

\author{
     Shikshya Shiwakoti\\ 
    \And Samuel Goldsmith\\ 
    \And  Ujjwal Pandit\\
    }
    
\begin{document}
\maketitle
\begin{abstract}
In today's information world, access to scientific publications has never been easier. At the same time, filtering through the mega corpus of information for research has never been more difficult. Graph Neural Networks (GNNs) and graph attention mechanisms have proven very useful in searching through large databases of information, especially when paired with modern large language models. In this paper, we propose our Attention Based Subgraph Retriever: a GNN-as-retriever model that uses attention pruning to pass a refined subgraph to an LLM for advanced knowledge reasoning.
\end{abstract}

\section{Introduction}
The internet is constantly growing, and with it, published research has never been more accessible. As more and more papers are released, finding relevant material grows exponentially more difficult. Naturally, growing alongside this difficulty is the need to create a streamlined way to find relevant and credible papers while leveraging their relations. In recent years, Artificial Intelligence has proven to be a powerful and consistent tool to address complex problems most humans would take months, even years, to complete. Large Language Models (LLMs) in particular have proven to be very promising, leveraging natural language to perform these complex tasks, including information retrieval. However, LLMs are prone to hallucination, and struggle with providing information that they were not trained on.\\
Recent years have shown retrieval-augmented generation, or RAG, is an excellent way to address these hallucinations by giving LLMs tools to parse large databases for relevant information \citep{rag}. Its success has inspired the adaptation of RAG to utilize the power of knowledge graphs \citep{graphrag}. GraphRAG has been shown to be quite promising, as it provides not just semantic context of information contained in these graphs, but also relational and structural information. Through leveraging knowledge graphs, it is possible to find information through relations between nodes; for example, relevant papers to one's research through analyzing similar papers and their citations, authors, and publication venues.
Graph Neural Networks (GNNs) can be used to harness the full power of these knowledge graphs when used as information retrievers for GraphRAG \citep{gnnrag}. GNN-RAG uses the GNN as a method to produce candidate context from a dense knowledge graph for an LLM to use. Taking inspiration from prior work, our project utilizes the power of homogeneous knowledge graph data and GNNs to generate citation recommendations given a candidate seed paper, which then gets reranked by an LLM agent. 

\section{Background}
\subsection{GNNs for Information Retrieval}
As previously mentioned, GNNs have proven to be beneficial for information retrieval. \citet{gnnrag} proposed GNN-RAG as their approach to leverage the power of knowledge graphs for question-answering tasks. The GNN reasons over a dense subgraph to provide candidate answers to a given question. This information is then given to an LLM, which reasons over the paths extracted from the knowledge graph to answer a given question. This approach was shown to be quite successful, further cementing the efficacy of GNNs for information retrieval. However, this approach fails to utilize the power of attention mechanisms that have become prevalent in modern GNNs.
\subsection{Attention and Pooling in GNNs}
\begin{figure}[h]
    \centering
    \includegraphics[width=0.9\linewidth]{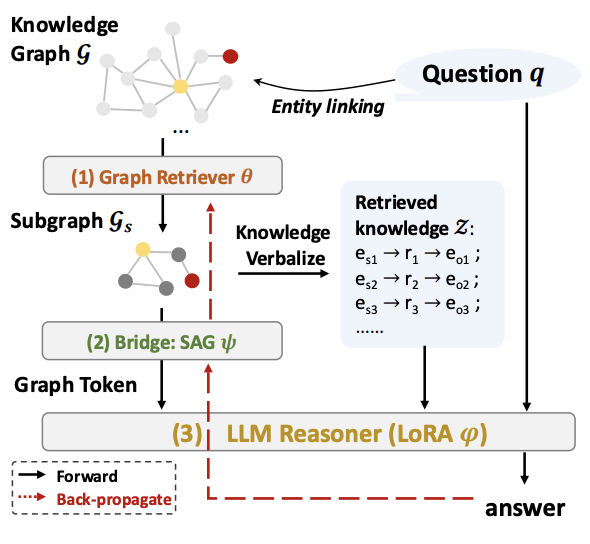}
    \caption{The framework of GRIL}
    \label{fig:gril}
\end{figure}

\begin{figure}[h]
    \centering
    \includegraphics[width=0.9\linewidth]{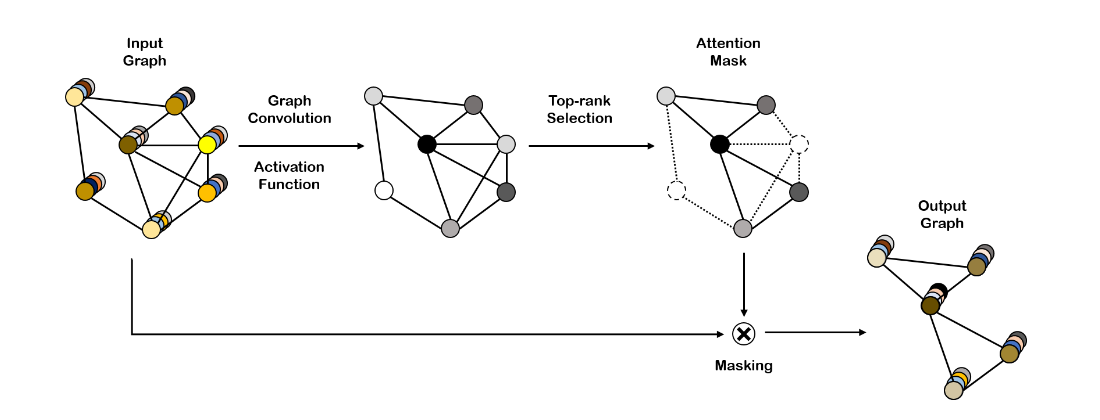}
    \caption{Outline of the SAGPool attention pruning mechanism.}
    \label{fig:sag}
\end{figure}

\begin{figure}[h]
    \centering
    \includegraphics[width=0.9\linewidth]{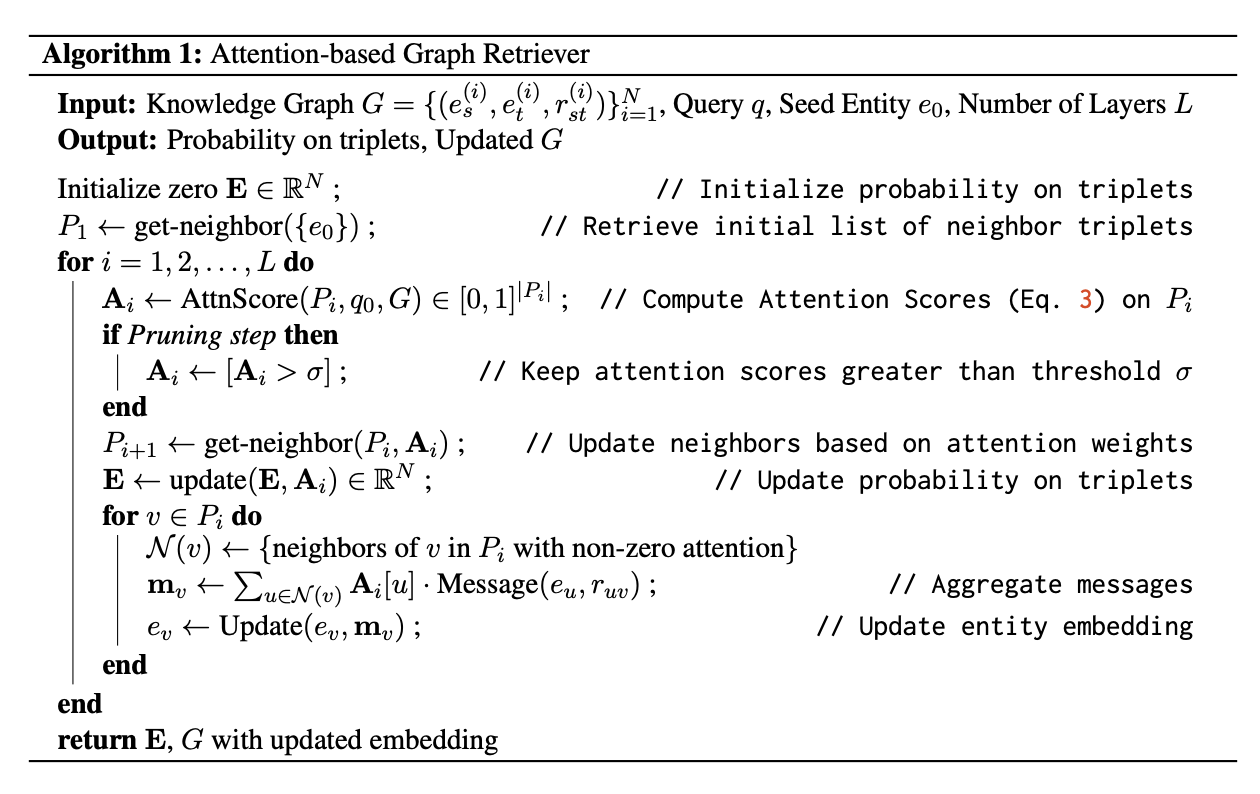}
    \caption{Algorithm of Attention-based Graph Retriever. It takes in the graph, query, seed, and layer count (number of hops) as input. It starts by initializing the frontier $P_1$ to the neighboring nodes of the seed. Then it computes attention scores by convolving over the subgraph in relation to the query. The subgraph then gets pruned where attention scores are below the desired hyperparameter sigma via SAGPool. This repeats the layer count amount of times, resulting in a subgraph with a max radius of L.}
    \label{fig:algorithm}
\end{figure}

Convolutions in neural networks are a well-known powerful concept that leverages spatial information from data. While originally designed for images, they have since been expanded into GNNs.\citet{gat} take this one step further in their graph-attention network (GAT) convolution layers by including attention mechanisms. Putting aside the fact that these attention mechanisms significantly reduce the computational cost of GNN convolutions and their lack of scalability, they also permit for advanced pruning methods where less-important features of neighboring nodes can be ignored. 
\cite{sagpool} proposes their Self-Attention Graph Pooling (SAGPool) method, which takes advantage of graph convolutional layers that produce attention scores that does just this- nodes and edges with attention scores less than a specified threshold get pruned from the outputted subgraph, as seen in figure \ref{fig:sag}. \citet{gril} utilizes the power of GNNs and SAGPool to create their novel Graph Retrieval-Integrated Learning (GRIL) framework. The GRIL pipeline as shown in Figure \ref{fig:gril} starts by using entity linking to get a query based seed node for a graph retriever. Using SAGPool, the graph retriever prunes off nodes from the subgraph output, which then gets expanded by the next graph convolution. This repeats for however many multihops are desired, as shown by the algorithm in figure \ref{fig:algorithm}. The output of this graph retriever is then passed to an LLM as a reasoner for Q\&A tasks.

\section{Approach}
Taking inspiration from prior work, we have built a GNN-as-retriever that uses attention scores and pruning mechanisms to recommend citations given a candidate paper.

\subsection{Dataset Overview and Pre-processing}
The Microsoft Academic Graph (MAG) dataset \citep{mag_dataset} is a collection of scientific publications, with data organized by authors, citations, publication dates, and publication venues, among other categories. It is a powerful dataset that has been used to train several of Microsoft's products, such as Cortana, Bing, and Word. For the purposes of our project, we will be using a subset of MAG\footnote{https://github.com/QianWangWPI/Released-Microsoft-dataset} to power our model.\\

Pre-processing was a challenging part of this work. The dataset consists of 42,000 training samples and 9,000 samples each for validation and testing.
The raw data is originally structured in JSON format. When converted into a table for pre-processing, the dataset contains the following columns: \\
\texttt{publication\_ID}, \texttt{Citations}, \texttt{pubDate}, \texttt{language}, \texttt{title}, \texttt{journal}, \texttt{abstract}, \texttt{keywords}, \texttt{authors}, \texttt{venue}, and \texttt{doi}.

The \texttt{Citations} column presented several challenges:
\begin{itemize}[noitemsep, topsep=0pt]
    \item Citations are mostly stored as lists of strings, but some entries are integers.
    \item Some entries contain missing values (\texttt{NaN}).
    \item A few publication IDs appear multiple times within the same \texttt{Citations} list.
\end{itemize}

The \texttt{pubDate} column has inconsistent formatting, too. Some dates are partial, in the format \texttt{yyyy-MMM} (e.g., 2008 Sep), while others represent a range of months, such as 2007 Mar-Apr. The dataset contains academic records from 1980 to 2014.

The \texttt{Authors} column is stored as a dictionary containing \texttt{name}, \texttt{id}, and \texttt{org}. However, not all records are consistent; most of the author names are reported as hashes, some are not hashed, and some are null. Some authors are reported with both names and ids, some are missing either or. The \texttt{Venue} column is structured similarly, with \texttt{name} and \texttt{id}, but most entries are missing \texttt{id}.

To create a homogeneous graph where only paper nodes are required, we concatenated the \texttt{title}, \texttt{abstract}, \texttt{keywords}, and \texttt{doi} columns to generate 384-dimensional embeddings. These embeddings were computed using the \texttt{all-MiniLM-L6-v2} sentence transformer model.

The resulting graph statistics are as follows:
\begin{itemize}[noitemsep, topsep=0pt]
    \item Training graph: 41,831 nodes and 3,401 edges
    \item Test graph: 8,967 nodes and 170 edges
    \item Validation graph: 8,970 nodes and 162 edges
\end{itemize}

\subsection{Implementation Specifications}
To create our GNN, we utilized the Pytorch and PyG libraries.
\subsection{Graph Retriever Workflow }
We have adapted a similar approach as used by \citet{gril}. First, we retrieved the seed node using semantic similarity between the query and the documents. This seed node is then passed into our graph retriever, which uses GATConv layers to produce attention scores for each node. While we do not use SAGPool, we use a similar attention pruning mechanism that removes less valuable nodes from the subgraph. We then expand the radius of the subgraph by 1, recalculate attention scores, and repeat this process for a specified number of hops.

At this stage, we obtain a graph of altered embeddings, which is then passed through the decoder to produce the final usable subgraph derived by the model. 

\subsection{Incorporating LLM-based Reranking}
In addition to Graph-based retrieval, we incorporated a large languae model(LLM) to re-rank the candidate papers retrieved from the subgraph. We used the \textit{'meta-llama/Meta-Llama-3-8B-Instruct'} model, quantized to 8-bit for efficient inference on GPU. 
\subsection{Evaluation Metrics}
To assess the performance of our graph-based citation retrieval system, we employed four standard evaluation metrics, all calculated at $k=10$ (top 10 results).

\textbf{Recall@k} measures the model’s ability to retrieve all relevant citations. Formally, it is defined as the proportion of relevant citations found within the top $k$ results:

\[
\text{Recall@k} = \frac{\text{No. of relevant citations in top } k}{\text{Total No. of relevant citations}}
\]

\textbf{Precision@k} evaluates the accuracy of the retrieved citations, defined as the fraction of citations in the top $k$ results that are truly relevant:

\[
\text{Precision@k} = \frac{\text{No. of relevant citations in top } k}{k}
\]

While Recall@k and Precision@k quantify the presence of relevant citations, they do not capture the quality of the ranking. To assess ranking quality, we used the following metrics:

\textbf{Mean Reciprocal Rank (MRR)} evaluates how early the first relevant citation appears in the ranked list. It is defined as the average reciprocal rank of the first relevant citation across all queries:

\[
\text{MRR} = \frac{1}{|Q|} \sum_{i=1}^{|Q|} \frac{1}{\text{rank}_i}
\]

where $\text{rank}_i$ is the position of the first relevant citation for query $i$. MRR emphasizes retrieving at least one relevant citation as early as possible.

\textbf{Normalized Discounted Cumulative Gain (nDCG@k)} measures the overall quality of the ranking, rewarding models for placing the most relevant citations higher in the list. It is computed as:

\[
\begin{aligned}
\text{nDCG@k} &= \frac{\text{DCG@k}}{\text{IDCG@k}}, \\
\text{DCG@k} &= \sum_{i=1}^{k} \frac{2^{\text{rel}_i} - 1}{\log_2(i+1)}
\end{aligned}
\]

where $\text{rel}_i$ represents the relevance of the $i$-th retrieved citation, and IDCG@k is the ideal DCG for the query. 

\section{Experimental Setup}
All experiments were conducted on a Google Colab Pro instance equipped with an NVIDIA L4 GPU. This environment was used for training, subgraph retrieval, and evaluation of all model configurations.
\paragraph{Model Configuration}
We employ the \textit{Attention Based Subgraph Retriever} architecture, which consists of three \texttt{GATConv} layers with ELU activations. Subgraph expansion is guided by attention scores computed by a lightweight attention network, parameterized by a pruning threshold $\sigma$. The maximum retrieval depth is determined by the hop parameter $L$.

\paragraph{Seed Node Initialization}
During training, a seed node was selected randomly from the training graph for each query to simulate a retrieval starting point. During evaluation, seed nodes were identified using a semantic similarity retriever based on SentenceTransformer embeddings.

\paragraph{Text Embeddings}
Text embeddings are generated using the \texttt{all-MiniLM-L6-v2} SentenceTransformer model. 

\section{Results}
\subsection{Evaluation on Test Subset}

To validate the performance of our graph-based citation retrieval system, we evaluated our model on a 1,000-paper test subset using four standard metrics: Precision@10, Recall@10, MRR and nDCG@10, all calculated for the top 10 results ($k=10$).

We compared our system against the following baselines:

\begin{itemize}
    \item \textbf{BM25:} A traditional information retrieval approach that uses TF-IDF and document length to find exact matches between query and corpus.
    \item \textbf{SBERT:} A dense embedding approach using the sentence transformer model \texttt{all-MiniLM-L6-v2} to compute cosine similarity between query and document embeddings.
    \item \textbf{Hybrid:} A combination of BM25 and SBERT scores: $\alpha \cdot \text{BM25} + (1-\alpha) \cdot \text{SBERT}$. We found that $\alpha=0.5$ (equal weighting) achieved the best performance.
\end{itemize}
The results of our evaluation are summarized in Table~\ref{tab:test_results}, which reports Precision@10, Recall@10, MRR, and nDCG@10 for our model and the baseline approaches.
As shown in the Table~\ref{tab:test_results}, our graph-based model, \textbf{Attention Based Subgraph Retriever}, achieves lower performance compared to the BM25, SBERT, and hybrid BM25+SBERT baselines on this 1,000-paper test subset. While the model demonstrates the potential of graph-based subgraph retrieval, further tuning and improvements are necessary to match or surpass the performance of established retrieval methods.

\begin{table*}[h]
\centering
\resizebox{\textwidth}{!}{%
\begin{tabular}{|c|l|c|c|c|c|}
\hline
\textbf{Category} & \textbf{Approach} & \textbf{Recall@10} & \textbf{Precision@10} & \textbf{MRR} & \textbf{nDCG@10} \\
\hline
Traditional IR & BM25 & 33.76\% & 4.10\% & 14.52\% & 17.25\% \\
Dense Embeddings & SBERT & 32.47\% & 3.97\% & 12.33\% & 15.73\% \\
Hybrid & \makecell[l]{$\alpha \cdot$BM25 + (1-$\alpha$) $\cdot$ SBERT \\ $\alpha=0.5$} & 47.43\% & 4.73\% & 15.85\% & 19.32\% \\
Our Model & Attention based subgraph retriever & 0.24\% & 0.26\% & 1.77\% & 1.38\%\\
Our Model+LLM Reasoning & Attention based subgraph retriever+LLM reranking & 0.3\% & 0.23\% & 0.03\% & 0.00\%\\
\hline
\end{tabular}%
}
\caption{Comparison of different retrieval approaches on the 1,000-paper test subset. Our Model+LLM Reasoning model was tested on only 100 papers. Metrics are computed for the top 10 retrieved results. Our graph-based model currently underperforms the baseline approaches.}
\label{tab:test_results}
\end{table*}

\subsection{Evaluation with LLM-based Re-ranking}

We evaluated our \texttt{AttentionRetrieverGAT} model on a 100-query subset of the test set, using LLM-based re-ranking for the top 10 retrieved nodes. 

The results are included in the Table~\ref{tab:test_results}.

As seen, the LLM re-ranking did not significantly improve retrieval quality, likely due to limited subgraph context and semantic complexity of the queries.

\subsection{Addressing Research Questions}
In this work, we aimed to answer two primary research questions:

\begin{enumerate}
    \item \textbf{Can we prompt an information retrieval model to recommend citations given a paper?} \\
    Yes. Our pipeline successfully processes a query, uses semantic similarity to identify a seed node, and employs the GNN-based Attention Retriever to identify the most relevant citations for recommendation.

    \item \textbf{Can we obtain the graph information retrieval model's reasoning when selecting documents for a given query?} \\
    Yes. The highly relevant papers retrieved from the subgraph are verbalized as knowledge triplets. This structural and relational information is passed to the LLM reasoner as context, enabling it to output reasoning alongside the recommendation.
\end{enumerate}

\section{Discussion and Challenges}
We faced three major challenges during the completion of this project. First, the pre-processing of the dataset was time-consuming, as aforementioned earlier. 

The second major challenge was the incompatibility of PyTorch's \texttt{SAGPool} layer with heterogeneous data. There were no credible sources providing guidance on this issue, and even implementing a custom \texttt{SAGPool} layer proved difficult under these conditions. 

The third, and perhaps the most difficult, challenge, driven by time constraints, was to pivot from a heterogeneous to a homogeneous graph midway through the project.

\section{Limitations}
Our approach utilizes only homogeneous data extracted from the dataset rather than the full, heterogeneous graph. While we had originally set out to use heterogeneous data, multiple setbacks required us to pivot to homogeneous data. Additionally, the dataset used is outdated, and our work may not scale to more modern academic publications. We also assume a static knowledge graph, and as such our work cannot be implemented without accommodating a dynamically growing/shrinking database.
\section{Future Work}
Future work should investigate utilizing the power of heterogeneous graphs. As they are quite powerful for representing complex relationships, modifying our approach to use heterogeneous data may prove useful. Additionally, the full MAG dataset should be considered. The subset used in this work was very sparse, resulting in lower performance than more dense knowledge graphs. Further work could also explore the robustness of this approach to erroneous edges in data, and explore modifying the model to handle a dynamically changing knowledge graph that gets added to and pruned regularly.

\bibliography{custom}

\begin{thebibliography}{7}
\providecommand{\natexlab}[1]{#1}

\bibitem[{Chen et~al.(2025)Chen, Zhang, Yun, Mottini, Ying, Song, Ioannidis, Li, and Cui}]{gril}
Jialin Chen, Houyu Zhang, Seongjun Yun, Alejandro Mottini, Rex Ying, Xiang Song, Vassilis~N. Ioannidis, Zheng Li, and Qingjun Cui. 2025.
\newblock \href {https://arxiv.org/abs/2509.16502} {Gril: Knowledge graph retrieval-integrated learning with large language models}.
\newblock \emph{Preprint}, arXiv:2509.16502.

\bibitem[{Han et~al.(2025)Han, Wang, Shomer, Guo, Ding, Lei, Halappanavar, Rossi, Mukherjee, Tang, He, Hua, Long, Zhao, Shah, Javari, Xia, and Tang}]{graphrag}
Haoyu Han, Yu~Wang, Harry Shomer, Kai Guo, Jiayuan Ding, Yongjia Lei, Mahantesh Halappanavar, Ryan~A. Rossi, Subhabrata Mukherjee, Xianfeng Tang, Qi~He, Zhigang Hua, Bo~Long, Tong Zhao, Neil Shah, Amin Javari, Yinglong Xia, and Jiliang Tang. 2025.
\newblock \href {https://arxiv.org/abs/2501.00309} {Retrieval-augmented generation with graphs (graphrag)}.
\newblock \emph{Preprint}, arXiv:2501.00309.

\bibitem[{Lee et~al.(2019)Lee, Lee, and Kang}]{sagpool}
Junhyun Lee, Inyeop Lee, and Jaewoo Kang. 2019.
\newblock \href {https://arxiv.org/abs/1904.08082} {Self-attention graph pooling}.
\newblock \emph{Preprint}, arXiv:1904.08082.

\bibitem[{Lewis et~al.(2021)Lewis, Perez, Piktus, Petroni, Karpukhin, Goyal, Küttler, Lewis, tau Yih, Rocktäschel, Riedel, and Kiela}]{rag}
Patrick Lewis, Ethan Perez, Aleksandra Piktus, Fabio Petroni, Vladimir Karpukhin, Naman Goyal, Heinrich Küttler, Mike Lewis, Wen tau Yih, Tim Rocktäschel, Sebastian Riedel, and Douwe Kiela. 2021.
\newblock \href {https://arxiv.org/abs/2005.11401} {Retrieval-augmented generation for knowledge-intensive nlp tasks}.
\newblock \emph{Preprint}, arXiv:2005.11401.

\bibitem[{Mavromatis and Karypis(2024)}]{gnnrag}
Costas Mavromatis and George Karypis. 2024.
\newblock \href {https://arxiv.org/abs/2405.20139} {Gnn-rag: Graph neural retrieval for large language model reasoning}.
\newblock \emph{Preprint}, arXiv:2405.20139.

\bibitem[{Veličković et~al.(2018)Veličković, Cucurull, Casanova, Romero, Liò, and Bengio}]{gat}
Petar Veličković, Guillem Cucurull, Arantxa Casanova, Adriana Romero, Pietro Liò, and Yoshua Bengio. 2018.
\newblock \href {https://arxiv.org/abs/1710.10903} {Graph attention networks}.
\newblock \emph{Preprint}, arXiv:1710.10903.

\bibitem[{Wang()}]{mag_dataset}
Qian Wang.
\newblock Released microsoft academic graph dataset.
\newblock \url{https://github.com/QianWangWPI/Released-Microsoft-dataset}.

\end{thebibliography}

\appendix



\end{document}